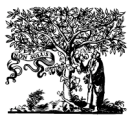
ELSEVIER

**PHYSICA C**

# Chemical trends of superconducting properties in pyrochlore oxides

Zenji Hiroi,[a,*] Jun-Ichi Yamaura,[a] Shigeki Yonezawa,[a] Hisatomo Harima[b]

[a]*Institute for Solid State Physics, University of Tokyo, Kashiwa, Chiba 277-8581, Japan*

[b]*Department of Physics, Kobe University, Kobe, Hyogo 657-8501, Japan*

**Abstract**

Chemical trends of fundamental superconducting parameters and normal-state properties are described for a family of pyrochlore oxide superconductors. Particularly, the change of $T_c$ from 1.0 K for $\alpha$-pyrochlore $Cd_2Re_2O_7$ to 3.3 K ($A$ = Cs), 6.3 K (Rb), and 9.6 K (K) for $\beta$-pyrochlore $AOs_2O_6$ is discussed on the basis of the conventional BCS scheme. Enhanced $T_c$ and anomalous features observed for $KOs_2O_6$ are ascribed to low-energy phonons probably coming from the rattling of the K cations.

## 1. Introduction

Twenty years have already passed after the discovery of Cu-based oxide superconductors in 1986 [1], and many related superconductors have been synthesized successfully. Consequently, it is getting more and more difficult to find a new one. On the other hand, the search for non-Cu-based oxide superconductors has been extended during the last decade, which mainly aims to understand the role of electron correlations in the mechanism of superconductivity or to search for a novel pairing mechanism, hopefully to reach a higher $T_c$.

An interesting example is a family of pyrochlore oxide superconductors. The first discovered is $\alpha$-pyrochlore $Cd_2Re_2O_7$ with $T_c$ = 1.0 K [2] and the second $\beta$-pyrochlore $AOs_2O_6$ with $T_c$ = 3.3, 6.3, and 9.6 K for $A$ = Cs [3], Rb [4-6], and K [7], respectively. They crystallize in the cubic pyrochlore structure of the same space group of $Fd$-$3m$ [8] and commonly possess a 3D skeleton made of $ReO_6$ or $OsO_6$ octahedra, as illustrated in Fig. 1. The difference between the two types comes from the fact that the O' atom at the 8$b$ site in the $\alpha$-pyrochlore $Cd_2Re_2O_6O'$ is replaced by the $A$ atom in the $\beta$-pyrochlore $AOs_2O_6$ [9, 10]. Moreover, the Cd 16$d$ site in the former is vacant in the latter. This means, in other words, that a relatively large $CdO_4$ tetrahedral unit is replaced by a single $A$ atom which must suffer a large size mismatch and can rattle in an oversized atomic cage. It was pointed out that this mismatch causes a peculiar anharmonic vibration, particularly for the smallest K atom [10-12].

The electronic structures of $\alpha$-$Cd_2Re_2O_7$ and $\beta$-$AOs_2O_6$ have been calculated by first-principle density-functional methods, which reveal that a metallic conduction occurs in the (Re, Os)-O network [12-16]: electronic states near the Fermi level originate from transition metal 5d and O 2p orbitals. Although the overall shape of the density of state (DOS) is similar for the two compounds, the difference in band filling results in different properties; $Re^{5+}$ for $\alpha$-$Cd_2Re_2O_7$ has two 5d electrons, while $Os^{5.5+}$ for $\beta$-$AOs_2O_6$ has two and a half. It is to be noted that a related $\alpha$-pyrochlore $Cd_2Os_2O_7$ with $Os^{5+}$ ($5d^3$) exhibits a metal-to-insulator transition at 230 K [17, 18].

The mechanism of superconductivity for the pyrochlore oxides has been studied extensively so far. Most of data obtained for $Cd_2Re_2O_7$ have revealed that it is a weak-coupling BCS-type superconductor [19, 20]. Experimental efforts to elucidate the nature of the superconductivity of the $\beta$-pyrochlores are in progress [11, 21-34]. Interestingly to be compared with $Cd_2Re_2O_7$, there are several findings which suggest unconventional features for $AOs_2O_6$. For example, resistivity exhibits an anomalous concave-downward curvature, indicating an unusual scattering process involved in the normal state [7]. Moreover, NMR experiments by Arai *et al.* showed a tiny coherence peak in the relaxation rate below $T_c$ for $RbOs_2O_6$, while no peaks were found for $KOs_2O_6$ [24]. This is in strong contrast to the result for $Cd_2Re_2O_7$ in which a huge coherence peak was detected [20]. In this paper, we consider systematic variations of various properties over the series, especially of $T_c$, in order to get meaningful insights on the mechanism of superconductivity and the source of the unconventional properties. The results of band structure calculations are also presented to understand the chemical trends.

---

* Corresponding author. Tel.: +81-47136-3445 ; fax: +81-47136-3446 ; e-mail: hiroi@issp.u-tokyo.ac.jp.

## 2. Experimental

Samples were prepared as reported previously [3, 4, 7, 9]. Single crystals of 1 mm size were obtained for $Cd_2Re_2O_7$ and $KOs_2O_6$, while only polycrystalline samples were available for $CsOs_2O_6$ and $RbOs_2O_6$, which caused a certain ambiguity for data obtained. The electrical resistivity and specific heat were measured in a Quantum Design physical property measurement system.

## 3. Results and discussion

### 3.1. Specific heat

Figure 2 compares specific heat data for the 4 compounds [11, 19, 22]. The superconducting transition is sharp for single crystals of $Cd_2Re_2O_7$ and $KOs_2O_6$, while is relatively broad for polycrystalline samples of $CsOs_2O_6$ and $RbOs_2O_6$. The magnitude of the jump at $T_c$, $\Delta C/\gamma T_c$, is 1.15 and 2.83 for $Cd_2Re_2O_7$ and $KOs_2O_6$, respectively, indicating that the latter is an extremely strong-coupling superconductor.

It is to be noted in Fig. 2 that the magnitude of specific heat above $T_c$ is very different among the 4 compounds. Compared at 10 K, for example, it is a factor of 8 larger for $KOs_2O_6$ than $Cd_2Re_2O_7$. This must be due to the difference in lattice contributions, not electronic ones, and is surprising because the lattice specific heat of such an identical or closely related structures should be similar. In the case of $Cd_2Re_2O_7$, the electronic and lattice parts could be reasonably separated as $C = \gamma T + \beta T^3$, assuming a Debye-type phonon at low temperature: $\gamma = 30.2$ mJ $K^{-2}$ $mol^{-1}$ and $\beta = 0.222$ mJ $K^{-4}$ $mol^{-1}$ that yields the Debye temperature $\Theta_D = 460$ K [19]. In contrast, it was found that the temperature dependence of the normal-state specific heat for the Cs and Rb compounds is unusually expressed as $C = \gamma T + \beta T^5$ in a wide temperature range between 0.5 and 7 K [11]. Very recently, we found the same $T^5$ dependence for a single crystalline $KOs_2O_6$ [35].

Thus, in the $\beta$ pyrochlores, usual Debye phonons may be masked by other types of low-energy phonons or other excitations. Possibly related to this, high-temperature specific heat is greatly enhanced due to the contribution from Einstein-like phonons. By fitting the data, Einstein temperature $\Theta_E$ was obtained as $\Theta_E$ = 70, 61, and 40 K, for $A$ = Cs, Rb, and K, respectively [11]. These must correspond to the specific frequency of the rattling motion of the alkali ions. The second anomaly observed at $T_p$ = 7.5 K for the specific heat of the K compound was reported in our previous paper, which may indicate a structural transition associated with rattling K cations [21, 22]. The $\gamma$ value and the magnitude of the jump at $T_c$ are given in Table 2.

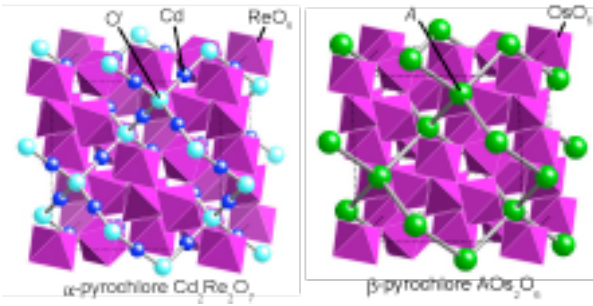

Fig. 1. Comparison of crystal structures between $\alpha$-pyrochlore $Cd_2Re_2O_7$ (left) and $\beta$-pyrochlore $AOs_2O_6$ (right). They possess the same skeleton made of $ReO_6$ or $OsO_6$. The difference comes from the fact that the O' atom in the former is replaced by the $A$ atom in the latter with the nearby Cd sites vacant.

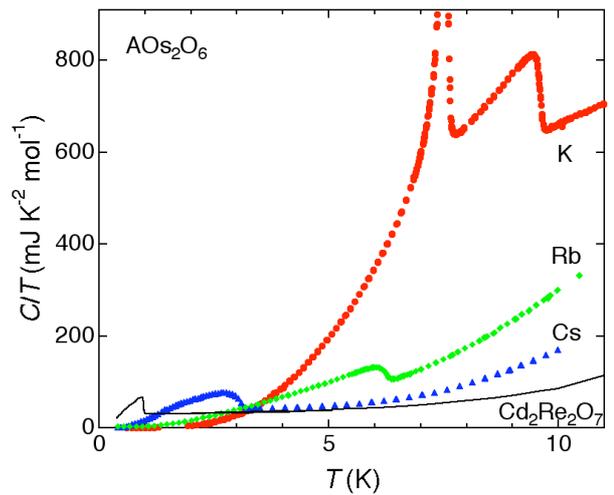

Fig. 2. Specific heat divided by temperature for $Cd_2Re_2O_7$ and $AOs_2O_6$. The data for $Cd_2Re_2O_7$ and $KOs_2O_6$ were obtained from single crystals and those for $CsOs_2O_6$ and $RbOs_2O_6$ from polycrystalline samples.

*3.2. Resistivity*

Figure 3 compares the temperature dependence of resistivity. Polycrystalline Cs and Rb samples exhibit a concave-downward resistivity at high temperature, followed by $T^2$ behavior at low temperature below a certain temperature $T^*$ of ~20 K and ~15 K, respectively. In contrast, the resistivity of the polycrystalline K sample shows a concave-downward curvature in a whole temperature range above $T_c$. Essentially the same temperature dependence is observed for a single crystalline K sample [22]. Figure 3(c) shows resistivity data measured at a magnetic field of 140 kOe for the single crystal of $KOs_2O_6$, where the $T_c$ is reduced to 5.2 K. Remarkably, a sudden drop in the normal-state resistivity is observed at $T_p$ = 7.5 K, where the specific heat exhibits a sharp peak (Fig. 2). Moreover, the temperature dependence below $T_p$ is proportional to $T^2$. These changes of the normal-state resistivity must imply that the scattering mechanism of carriers is substantially changed at $T_p$. Consequently, commonly observed in the resistivity of the $\beta$ pyrochlores is the change from high-temperature concave-downward to low-temperature $T^2$ behavior, which takes place as a crossover at $T^*$ for Cs and Rb, while as a phase transition at $T_p$ for K. The former temperature dependence suggests a strong electron-phonon interaction possibly ascribed to the rattling of $A$ cations, while the latter may indicate that electron-electron scattering dominates at low temperature, as suggested also by recent thermal conductivity measurements [34]. From a structural point of view, this implies that the rattling motion is frozen below $T^*$ for Cs and Rb, while it undergoes a cooperative phenomenon at $T_p$ for K due to a finite interaction between the rattlers [22].

In contrast, the resistivity of $\alpha$-pyrochlore $Cd_2Re_2O_7$ exhibits more conventional features at high temperature, as shown in Fig. 3(a). Moreover, $T^3$ behavior, not $T^2$, is observed at low temperature below 27 K, as shown in Fig. 3(c), which may be due to scattering by phonons on the cage.

One more remark on the single crystal of $KOs_2O_6$ is its small residual resistivity, $\rho_0 \sim 1~\mu\Omega$ cm, which was estimated by extrapolating the $T^2$ dependence to $T = 0$. Since the room temperature resistivity is ~300 $\mu\Omega$cm, the residual resistivity ratio (RRR) reaches 300, which is unusually large for transition metal oxides. The RRR of the $Cd_2Re_2O_7$ crystal shown in Fig. 3 is only 30. The residual resistivity is generally given as

$$\rho_0 = \frac{\hbar(3\pi^2)^{1/3}}{e^2 \ell n^{-2/3}},$$

where $l$ and $n$ are the mean-free-path and density of carriers, respectively. Since the carrier density has not yet been determined experimentally, we calculated it from the band structure calculations mentioned later; $n = 2.8 \times 10^{21}$ cm$^{-3}$.

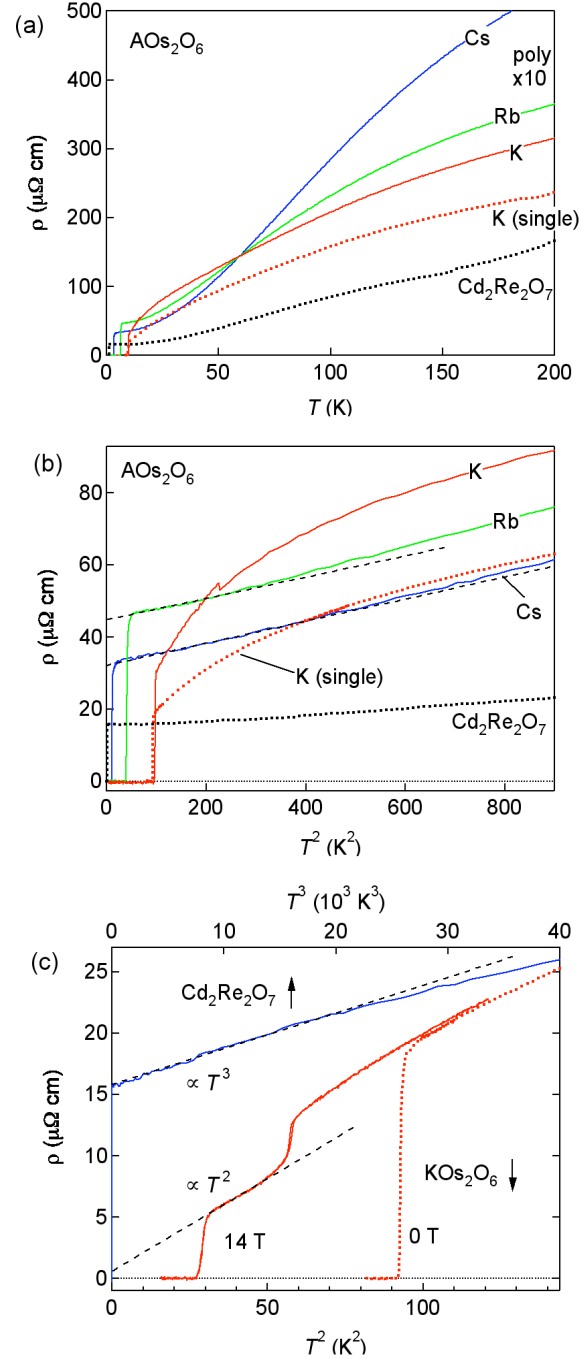

Fig. 3. Resistivity for polycrystalline $AOs_2O_6$ (solid lines) and single-crystalline $KOs_2O_6$ and $Cd_2Re_2O_7$ (dotted lines) plotted as a function of $T$ in a wide temperature range (a) and as a function of $T^2$ at low temperature below 30 K (b). In (c), resistivity of the single-crystalline $KOs_2O_6$ measured at a magnetic field of 14 T is plotted against $T^2$ (lower axis), and the resistivity of $Cd_2Re_2O_7$ is plotted against $T^3$ (upper axis). Broken straight lines are the guides to the eye.

Then, we obtain a large mean-free-path of 640 nm, which is comparable to the value of the most cleanest crystal of $Sr_2RuO_4$ [36]. Since the superconducting coherence length $\xi$ of $KOs_2O_6$ is 3.3 nm, it is definitely in the regime of ultra clean limit.

Table 1
Summary of superconducting and normal-state properties for $\alpha$-pyrochlore $Cd_2Re_2O_7$ and $\beta$-pyrochlore $AOs_2O_6$ of $A$= Cs, Rb, K

|   | $Cd_2Re_2O_7$ | $CsOs_2O_6$ | $RbOs_2O_6$ | $KOs_2O_6$ |
| --- | --- | --- | --- | --- |
| $T_c$ (K) | 1.0 | 3.3 | 6.3 | 9.6 |
| $\mu_0 H_{c2}$ (T) | 0.29 | ~3.3 | ~5.5 | 30.6 |
| $\xi$ (nm) | 34 | ~10 | ~7.8 | 3.3 |
| $\lambda$ (nm) | 460 | 400 | 234 | 270 |
| GL parameter | 14 | ~40 | ~30 | 82 |
| $\Delta C/\gamma T_c$ | 1.15 | ~1 | ~1 | 2.87 |
| $\gamma_{exp}$ (mJ $K^{-2}$ $mol^{-1}$) | 30.2 | ~40 | ~40 | 71 |
| $\gamma_{band}$ (mJ $K^{-2}$ $mol^{-1}$) | 11.5 | 11.0 | 10.2 | 9.6 |

Table 2
Summary of structural parameters for $\alpha$-pyrochlore $Cd_2Re_2O_7$ and $\beta$-pyrochlore $AOs_2O_6$ of $A$= Cs, Rb, K

|   | $Cd_2Re_2O_7$ | $CsOs_2O_6$ | $RbOs_2O_6$ | $KOs_2O_6$ |
| --- | --- | --- | --- | --- |
| $a$ (nm) | 1.02232 | 1.01525 | 1.01176 | 1.01065 |
| $x$ (O)[a] | 0.3137 | 0.3146 | 0.3180 | 0.3160 |
| $U_{iso}$ ($10^{-4}$ $nm^2$)[b] | 1.1 | 2.5 | 4.3 | 7.7 |
| Ionic radius (nm) | 0.095 | 0.167 | 0.152 | 0.138 |
| $\Theta_E$ (K) | - | 70 | 61 | ~40 |

[a] Atomic coordinate of 48f oxygen; ($x$ 0 0).

[b] Atomic displacement parameter for Cd or A.

### 3.3. Chemical trends

Figure 4 summarizes the evolution of several experimental parameters, $\gamma$, $\Delta C/\gamma T_c$, $H_{c2}$, $\Theta_E$ and $U_{iso}$, as a function of $T_c$. The $\gamma$ value is much larger in each case than the bare electronic value obtained from band structure calculations, which decreases slightly with increasing $T_c$, as reported previously [12, 16]. The enhancement factor $\gamma_{exp}/\gamma_{band}$ is 2.6 (Re), 3.6 (Cs), 3.9 (Rb), and 7.4 (K), which give large renormalization factors $\lambda$ in $\gamma_{exp}/\gamma_{band} = 1 + \lambda$. The relation between $\gamma$ and $T_c$ is discussed later. To be stressed here is the exceptionally large enhancement for K. This is also the case for other parameters, $\Delta C/\gamma T_c$ and $H_{c2}$, suggesting some additional effect must be present for K. The $H_p$ in Fig. 4(c) is the paramagnetic limit of the upper critical field simply calculated as $H_p = 1.85 T_c$. The $H_{c2}$ value is lower than $H_p$ except for the case of K. However, it was pointed out for K, by taking account of experimental normal-state magnetic susceptibility, that an actual $H_p$ is approximately 30 T, close to the experimental value [32, 34].

On the other hand, the structural parameters change systematically. The Einstein temperature decreases and the atomic displacement parameter of the A cations increases from Cs to K [9, 10]. This implies that the characteristic energy of rattling is lowered toward K with decreasing the size of the rattler and thus increasing the available space in a cage. Several structural parameters are listed in Table 2.

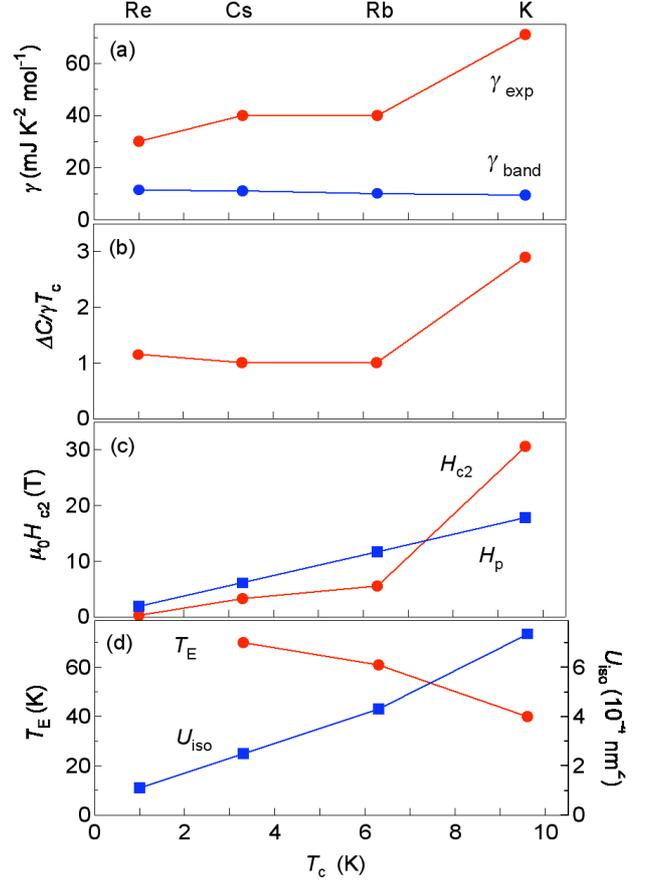

Fig.4. Evolution of Sommerfeld coefficient $\gamma$ (a), jump in specific heat at $T_c$, $\Delta C/\gamma T_c$ (b), upper critical field $H_{c2}$ (c), and Einstein temperature $\Theta_E$ and atomic displacement parameter $U_{iso}$ (d) as a function of $T_c$.

### 3.4. Bandstructure calculations

Electronic bandstructure calculations were carried out for $\alpha$-pyrochlores $Cd_2Re_2O_7$ and $Cd_2Os_2O_7$ as well as for $\beta$-pyrochlores $AOs_2O_6$. As reported already by other groups [12-16], these 5d pyrochlore oxides possess a strongly hybridized band made of Re/Os 5d and O 2p states near the Fermi level, which is a manifold of 12 bands with a total bandwidth of ~3 eV. Because of the large bandwidth, they may not be classified as strongly correlated electron systems in the ordinary sense. As shown in Fig. 5, the profile of the density-of-states (DOS) is similar among them, exhibiting many sharp peaks due to a large degeneracy coming from the high symmetry of the pyrochlore structure. The major difference between α and β pyrochlores comes from the fact that there is a finite contribution from the Cd 4d states to the $E_F$ state in α, while almost nothing from alkali metals in β.

The great variation of physical properties arises from band filling; $5d^2$, $5d^{2.5}$ and $5d^3$ for $Cd_2Re_2O_7$, $AOs_2O_6$, and $Cd_2Os_2O_7$. As shown in Fig. 5, the $E_F$ is located near a valley in $Cd_2Re_2O_7$, which results in a small Fermi surface around the zone center and a low carrier density [13]. In contrast, the $E_F$ happens to lie close to a sharp peak at the middle of the manifold for $Cd_2Os_2O_7$ [14], which causes a certain magnetic instability leading to a metal-to-insulator transition (the low-temperature phase is not completely insulating, because of a part of the Fermi surface may survive). Thus, electron correlations may play a role in $Cd_2Os_2O_7$ to some extent [18]. In contrast, β-pyrochlore $AOs_2O_6$ possesses a moderately large DOS between the two α pyrochlores, where one expects a moderately large electron correlations.

The evolution of the bandstructure for the β pyrochlores from Cs to K is unexpectedly subtle, as reported in previous study [12, 15, 16]. Because almost no contributions to the bandstructure arise from alkali metals in β, the bandstructure is roughly scaled to the lattice constant: the smaller the lattice, where larger hybridizations occur, the smaller the calculated DOS, as shown in Fig. 4. This is apparently different from the experimental observations. In order to explain the observed large enhancement of γ toward K, an additional effect must be taken into account. Here we point out that a significant change occurs for a pair of Fermi sheets over the series of compounds.

The calculated Fermi surface of β-$AOs_2O_6$ consists of a pair of closed sheets around the zone center and a hole-like sheet near the zone boundary. As shown in Fig. 6, the former Fermi surface possesses a characteristic shape: each sheet is not spherical but close to an octahedron with flat surfaces, suggesting an electronic instability due to Fermi surface nesting. Moreover, the two sheets are nearly homothetic. There is an interesting theoretical argument that such disconnected Fermi surfaces would suppress pair breaking and thus lead to an enhanced $T_c$ [37]. It is apparent from the cross section perpendicular to the [110] direction, as illustrated in Fig. 6, that the two sheets are nearly parallel to each other for K, while the outer sheet dents toward the inner sheet for Cs. Thus, nesting tendency must be enhanced from Cs to K. It is plausible that this nesting feature plays some role in the mass enhancement and also the superconductivity in the β pyrochlores.

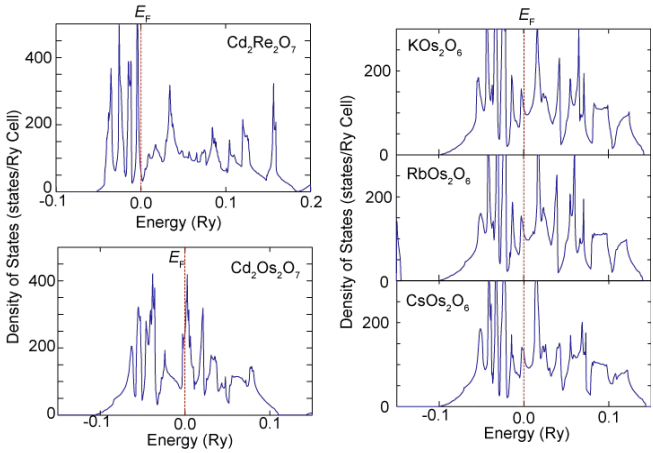

Fig. 5. Density of states near the Fermi level of the pyrochlore oxides.

### 3.5. Normal-state properties

Here we consider electron correlations and mass enhancements for the pyrochlore oxides. Figure 7 displays a Kadowaki-Woods plot where many metallic compounds are mapped in the space of the coefficient $A$ of the $T^2$ term in resistivity and the Sommerfeld coefficient γ [38, 39]. In the case of $KOs_2O_6$, the $A$ value of 0.143 μΩ cm $K^{-2}$ from Fig. 3(c) and the γ value of 71 mJ $K^{-2}$ $mol^{-1}$ yield $A/\gamma^2 = 2.8 \times 10^{-5}$, close to the universal value of $10^{-5}$, which implies that $KOs_2O_6$ is akin to the strongly correlated system. For Cs and Rb, we could not determine the value, because the $A$ value must be overestimated owing to the polycrystalline nature.

Another important parameter for the strongly correlated system is the Wilson ratio $R_W$ that is $\chi_s/\gamma$. In the case of $Cd_2Re_2O_7$, the magnetic susceptibility measurements gave $\chi_0 = 4.7 \times 10^{-4}$ emu/mol, which should include an orbital contribution of $3.2 \times 10^{-4}$ emu/mol, as determined from Cd NMR measurements [40]. Thus, the spin susceptibility $\chi_s = 1.5 \times 10^{-4}$ emu/mol, which yields $R_W = 0.34$. This small $R_W$ may be consistent with the absence of $T^2$ term in resistivity. On the other hand, the $\chi_0$ value of the present single crystalline $KOs_2O_6$ was $\chi_0 = 4.6 \times 10^{-4}$ emu/mol, similar as for $Cd_2Re_2O_7$. Assuming the same orbital contribution as for $Cd_2Re_2O_7$, $\chi_s = 1.4 \times 10^{-4}$ emu/mol, which gives a small $R_W$ of 0.14. Such a small values of $R_W$ generally implies strong electron-phonon interactions that must enhance only $\gamma$ and less importance of conventional electron correlations.

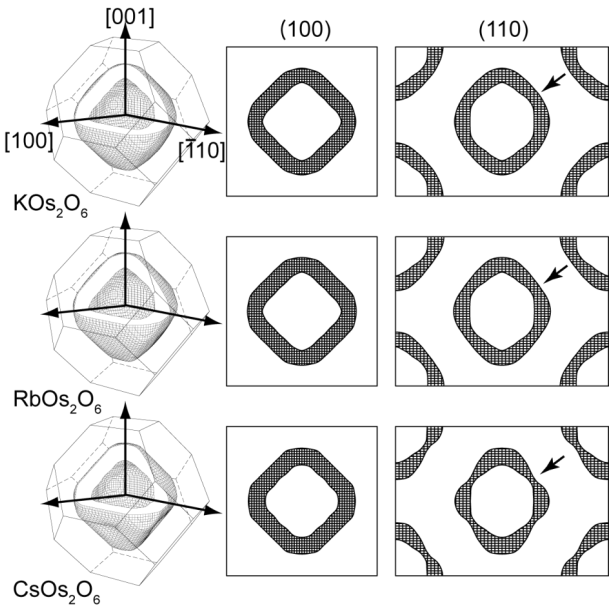

Fig. 6. Calculated Fermi surfaces for the three members of the β pyrochlores. (100) and (110) cross sections are also shown.

In the case of Cs and Rb, the $\chi_0$ value is similar to that of K, although there is an uncertainty from the sample quality. The $R_W$ value may be also small, ~0.25.

$$T_c = \omega_{ph} \exp\left(-\frac{1}{N(0)V}\right)$$

where $\omega_{ph}$ is a frequency of a relevant phonon, $N(0)$ the DOS at the Fermi level, and $V$ the pairing potential. Let us assume that $T_c$ depends only on the $N(0)$, keeping both the $\omega_{ph}$ and $V$ constant over the compounds. Then, the $T_c$ of one compound can be calculated from that of the other as

$$T_c' = \omega_{ph} \exp\left[\frac{N(0)}{N'(0)} \ln\left(\frac{T_c}{\omega_{ph}}\right)\right].$$

Provided $T_c = 1.0$ K and $\gamma = 30.2$ mJ K$^{-2}$ mol$^{-1}$ ($\gamma$ is proportional to $N(0)$) for $Cd_2Re_2O_7$ as a reference, one can estimate $T_c$'s for the others for a certain phonon energy. Figure 8(b) depicts the results for $\omega_{ph} = 460$ K, which is the Debye temperature of $Cd_2Re_2O_7$ [2], 300 K, and 50 K. Apparently, the two high-energy phonons can reproduce $T_c$ for Cs and Rb ($T_c' = 4.5$ K and 4.0 K for $\omega_{ph} = 460$ K and 300 K), while overestimate for K. On the other hand, the

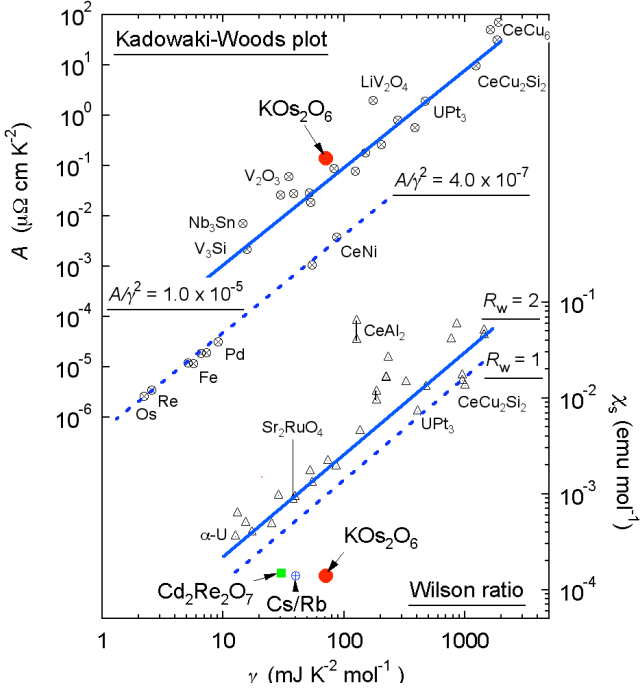

Fig. 7. Kadowaki-Woods plot and the Wilson ratio.

*3.6. Superconductivity*

The chemical trend of $T_c$ is now discussed in the framework of conventional BCS theory. $T_c$ is plotted as a function of the lattice constant $a$ in Fig. 8(a). As the lattice is compressed by the chemical pressure, the $T_c$ increases steeply toward K. This tendency is not compatible with a simple argument on a single-band model. However, it may not be surprising from the multi-band nature of these oxides. Saniz and Freeman discussed the trend of $T_c$ and estimated $T_c$ based on the McMillan-Allen-Dynes formalism [16]. They claimed that the incorporation of an electron-spin coupling is required to reproduce the experimental $T_c$.

We discuss in a simple way to deduce the relationship between $T_c$ and DOS, using an equation low-energy phonons of 50 K gives $T_c' = 2.6$ K for Cs/Rb and 9.5 K for K, in good agreement for K. Therefore, one identical phonon can not explain the change of $T_c$. It is plausible that a relevant phonon energy decreases for $KOs_2O_6$, which suppress the increase of $T_c$ due to the rise in the DOS.

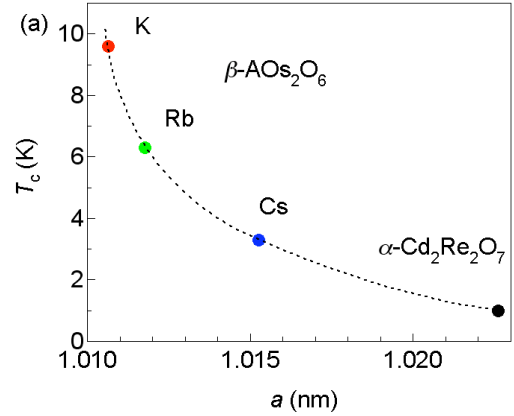

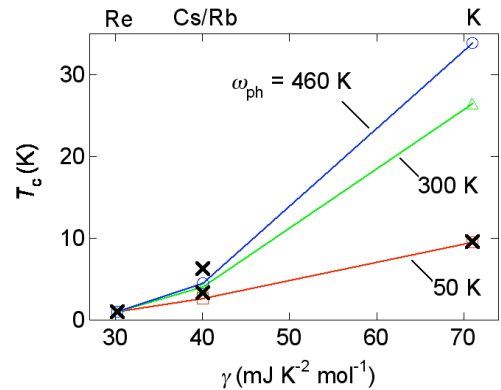

Fig. 8. (a) Variation of $T_c$ with the lattice constant $a$ for pyrochlore oxide superconductors. (b) Comparison of experimental and calculated $T_c$'s as a function of the Sommerfeld coefficient $\gamma$. Experimental $T_c$ values are shown by crosses, and those calculated by open marks for 3 cases of different phonons; $\omega_{ph} = 460$ K, 300 K, and 50 K.

The role of low energy phonons on the mechanism of superconductivity has been studied in many strong-coupling superconductors, such as Pb, A15 and Chevrel phase compounds. The correction for the BCS ratio, $\Delta C/\gamma T_c$, by the strong electron-phonon interactions was deduced in the following form [41];

$$\frac{\Delta C}{\gamma T_c} = 1.43\left[1 + 53\left(\frac{T_c}{\omega_{\log}}\right)^2 \ln\left(\frac{\omega_{\log}}{3T_c}\right)\right].$$

Using this equation, we estimate $\omega_{\log}$ to be 59 K for $KOs_2O_6$. Then, a coupling constant $\lambda$ can be obtained from the McMillan-Allen-Dynes equation [42];

$$T_c = \frac{\omega_{\log}}{1.2}\exp\left[\frac{-1.04(1+\lambda)}{\lambda - \mu^*(1+0.62\lambda)}\right],$$

where $\mu^*$ is the Coulomb coupling constant and was calculated to be 0.096, 0.093, and 0.091 for Cs, Rb, and K, respectively [16]. The $\lambda$ value for K is thus determined to be 2.38, which is much larger than that of typical strong-coupling superconductors like Pb ($\lambda = 1.55$, $\omega_{\log} = 50$ K) and $Mo_6Se_8$ ($\lambda = 1.27$, $\omega_{\log} = 70$ K) and is comparable with that of a Bi-Pb alloy ($\lambda = 2.1$, $\omega_{\log} = 45$ K) [42]. Therefore, $KOs_2O_6$ is an extremely strong-coupling superconductor.

It is reasonable to ascribe such a low-energy phonon to the rattling vibration in the case of the β pyrochlores. The observed unusual enhancements of $\gamma$ or $H_{c2}$ toward K must be related to the lowering of the characteristic energy of the rattling, as shown in Fig. 4. It would be an intriguing question how this rattling, which is an essentially anharmonic vibration almost localized in a cage, can mediate Cooper pairing in $KOs_2O_6$.

### 3.7. Pressure dependence of $T_c$

Pressure effects on $T_c$ for the β pyrochlores have been studied experimentally [26, 29, 43] and theoretically [16]. As expected from the dependence of $T_c$ on the lattice constant shown in Fig. 8(a), the $T_c$ initially increases with compressing the lattice by applying physical pressure. Then, it reaches a maximum value, gradually decreases with further increasing pressure, and is finally suppressed above a critical pressure [26]. Although the maximum value and the critical pressure depend on the *A* elements, a bell-shaped $T_c$ variation is commonly observed. Saniz and Freeman evaluated the pressure effects quantitatively and found that pressure effects on the overall electronic structures are rather small [16]. It was found, however, that $T_c$ increases with pressure in spite that both the DOS and $\lambda_{ep}$ decrease with pressure. This is mainly because $\mu^*$ as well as $\mu_{ep}$ (the electron-spin coupling constant), which are effective to reduce $T_c$, decrease more rapidly.

In order to get more insights on what happens under high pressure, we carried out structural study under high pressure up to 7 GPa and obtained volume compressibility for all the members of β pyrochlores. It is almost linear with pressure for Rb and K and tends to saturate for Cs. By fitting the data to the form, $\Delta V/V_0 = -\alpha P + \beta P^2$, we obtained that $\alpha = 0.00726$ $(GPa)^{-1}$ for K, $\alpha = 0.00777$ $(GPa)^{-1}$ for Rb, and $\alpha = 0.00756$ $(GPa)^{-1}$ and $\beta = 0.000198$ $(GPa)^{-2}$ for Cs. Using these values, the pressure dependence of $T_c$ reported in Ref. 26 is transformed to a volume dependence of $T_c$ as shown in Fig. 9. Interestingly, the three sets of data merge into a single straight line at low volume. Deviation from this universal line occurs at a different volume that increases from Cs to K. It seems that K takes the highest $T_c$ due to the fact that it keeps the universal line up to the largest volume. The initial increase of $T_c$ with volume may be attributed to the increase of DOS according to the bandstructure calculations. On the other hand, the decrease of $T_c$ above the maximum may be related to the decrease in the energy of relevant phonons. Further investigations are required to understand these features.

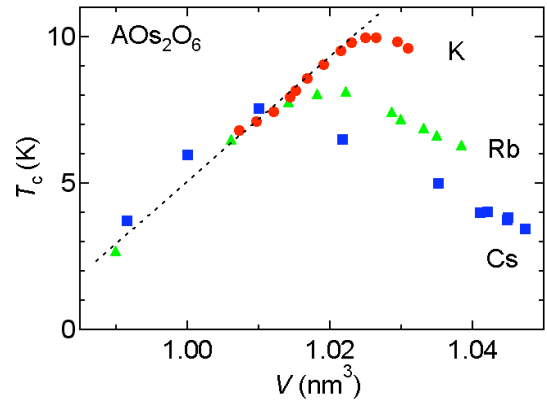

Fig. 9. Cell volume dependence of $T_c$

### 4. Concluding remarks

Finally, we add 4 points for the pyrochlore oxide superconductors to the well-known $T_c$-$\gamma$ diagram shown in Fig. 10. The β-pyrochlores, especially $KOs_2O_6$, exist close to the Chevrel phase family in the map, and are characterized as a new member of strong-coupling superconductors. It is suggested in the β pyrochlores, as in other strong-coupling superconductors, that low-energy phonons play a crucial role in the pairing mechanism, and that it comes from the rattling of the alkali metal ions. We think that future systematic study on the pyrochlore superconductors would extend the horizon for understanding the physics of strong-coupling superconductivity.

## Acknowledgments

We thank M. Takigawa and Y. Matsuda for enlightening discussions, and N. Takeshita and his coworkers for providing unpublished high-pressure data on a single crystal of $KOs_2O_6$. This research was supported by a Grant-in-Aid for Scientific Research B (16340101) provided by the Ministry of Education, Culture, Sports, Science and Technology, Japan.

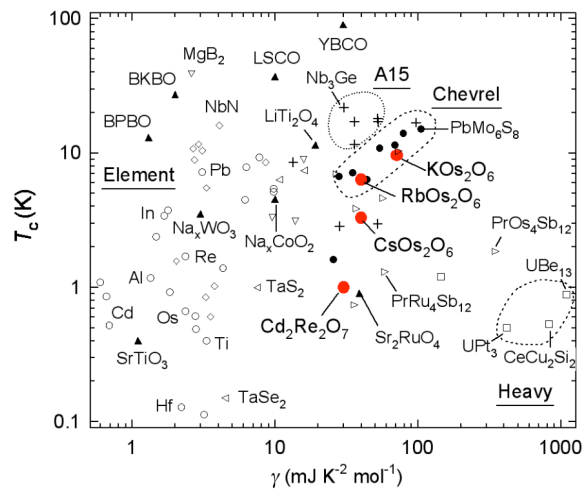

Fig. 10. Various superconductors mapped in the $T_c$-$\gamma$ diagram.